Research Article

# Experimental Evidence Supporting a New Osmosis Law & Theory Derived New Formula that Improves van't Hoff Osmotic Pressure Equation

Rongqing Xie,[a] Gaochao Lin,[*][b] and Hung-Chung Huang[*][c]

***Abstract***

*Experimental data were used to support a new concept "osmotic force" and a new osmotic law that can explain the osmotic process without the difficulties encountered with van't Hoff osmotic pressure theory. Derived new osmotic formula with curvilinear equation (via new osmotic law) overcomes the limitations and incompleteness of van't Hoff (linear) osmotic pressure equation, $\pi = \frac{n}{v}RT$, (for ideal dilute solution only). The application of this classical theory often resulted in contradiction regardless of miscellaneous explaining efforts. This is due to the lack of a scientific concept like "osmotic force" that we believe can elaborate the osmotic process. Via this new concept, the proposed new osmotic law and derived new osmotic pressure equation will greatly complete and improve the theoretical consistency within the scientific framework of osmosis.*

## 1. INTRODUCTION

## Osmotic force, osmotic law and their explanation on osmotic process

### 1.1 osmotic force

Osmotic force is the force that drives solvent to penetrate through the semi-permeable membrane (abbreviated as "membrane" in this article). Its value is equal to that pressure intensity (i.e., force per unit area) on the membrane times the osmotic effective area, or that pressure force on the membrane times the ratio of osmotic effective area to whole area of membrane.

The stress or force on the membrane consists of not only the hydraulic pressure, but also the atmospheric pressure. Although the atmospheric pressures are the same on both sides of membrane, their strengthening effects on osmotic forces are different due to different solution concentration on the opposite sides of the membrane due to different osmotic effective areas. In other words, the same amount of atmospheric pressure causes different increments of osmotic force subjected to different solution concentrations on different sides. The greater the difference between concentrations, the larger the gap between osmotic forces. Therefore, the atmospheric pressures on each side of membrane could not be ignored or offset.

The osmotic effective area is the area contacted by the solvent molecules. Although the whole area of membrane is contacted by both solvent and solute molecules, the area which is only contacted by solvent molecules accounts for osmotic effective area (the remaining area could be called "osmotic ineffective area"). Although the absolute positions of the effective and ineffective areas change instantaneously, the ratio between them is constant for a given concentration of solution. The absolute areas of the aforementioned two types are difficult to determine, but they could be derived from osmotic equilibrium when the osmotic forces of two sides of the membrane are the same under the condition that the effective area on the pure solvent side is 100% (as seen in Fig. 1 below). During the conversion process, effective area and ineffective area are presented by percentage occupying the whole area, so their sum is 1.

[*] Corresponding authors
[a] Dr. R. Xie
Department of Chemistry
Zhengzhou Normal University
6 Yingcai St., Huiji District, Zhengzhou, Henan, China 450044
E-mail: 1844168151@qq.com
[b] G. Lin
School of Civil and Environmental Engineering
UNSW (the University of New South Wales)
Sydney, NSW 2052, AUSTRALIA
E-mail: gaochao.lin@unsw.edu.au
[c] Dr. H.-C. Huang
Department of Biology
Jackson State University
1400 John R. Lynch Street, Jackson, MS 39217
E-mail: hung-chung.huang@jsums.edu; zoello.huang@gmail.com

From the above mentioned, relevant formulas can be derived:

(1) Osmotic force formula:

$$\mathrm{f} = P_X \cdot s$$

in which f is osmotic force, $P_X$ is the pressure intensity on membrane (i.e., force per unit area), and s is the osmotic effective area. Or in form of:

$$\mathrm{f} = P \cdot s_X$$

in which $P$ is the pressure force on the membrane, and $s_X$ is the percentage of osmotic effective area over total membrane area. Or in form of:

$$\mathrm{f} = P_X \cdot S \cdot (1 - k[C_i])$$

in which S is the total membrane area, $[C_i]$ is solute's molar concentration in solution, $k$ (related to solute concentration) is defined as the osmotic ineffective membrane area constant of ideal solution, and $(1 - k[C_i])$ is the percentage of osmotic effective area. $k$'s value is equal to the percentage of osmotic ineffective membrane area when activity coefficient is equal to 1 or the effective concentration is equal to 1 mol/L.

(2) net osmotic force formula:

$$\Delta\mathrm{f} = P_a \cdot s_a - P_b \cdot s_b$$

in which, $P_a$, $s_a$ and $P_b$, $s_b$ are the intensity of pressure and osmotic effective area on the a-side and b-side of membrane respectively. If $\Delta\mathrm{f} > 0$, the net osmosis happens from a-side to b-side; if $\Delta\mathrm{f} < 0$, the net osmosis happens from b-side to a-side.

(3) osmotic equilibrium formula:

$$P_a \cdot s_a = P_b \cdot s_b$$

at this moment, the osmosis is in dynamic equilibrium.

## 1.2 osmotic law

It is the process that solvent moves from the side with higher osmotic force to the other side with lower osmotic force until reaching the dynamic equilibrium in an osmotic system.

## 1.3 Explaining the osmotic process with osmotic law

The previous explanation about osmosis with classical osmotic pressure theory encountered conflict between ideal and non-ideal solutions. The shortage and incompleteness of van't Hoff equation(Kiil, 2003)(Prickett et al., 2008)(Rodenburg et al., 2017)(Lopez and Hall, 2020) encountered could be eliminated by utilization of proposed new osmotic law which is based on the new concept of "osmotic force".

In the initial stage of osmosis as seen in Fig. 1A, solvent (left side, or relatively dilute solution) and solution (right side, or relatively concentrated solution) have the same liquid level. Because the solvent side possesses higher osmotic force, the solvent then begins to move to solution side, which results in the increase of hydraulic pressure on the solution side. In other words, during this process, the osmotic force in solvent side gradually decreases due to declining pressure (if it were relatively dilute solution, the osmotic effective area decreases as well), while the osmotic force on solution side gradually increases due to increasing pressure strength from more waters passing to this side and more osmotic effective area formed due to solutes being able to distribute into more space on this side now (Fig. 1B). When the osmotic forces between both sides reach the same, the liquid level stops changing, and the system arrives at equilibrium state (Fig. 1C).

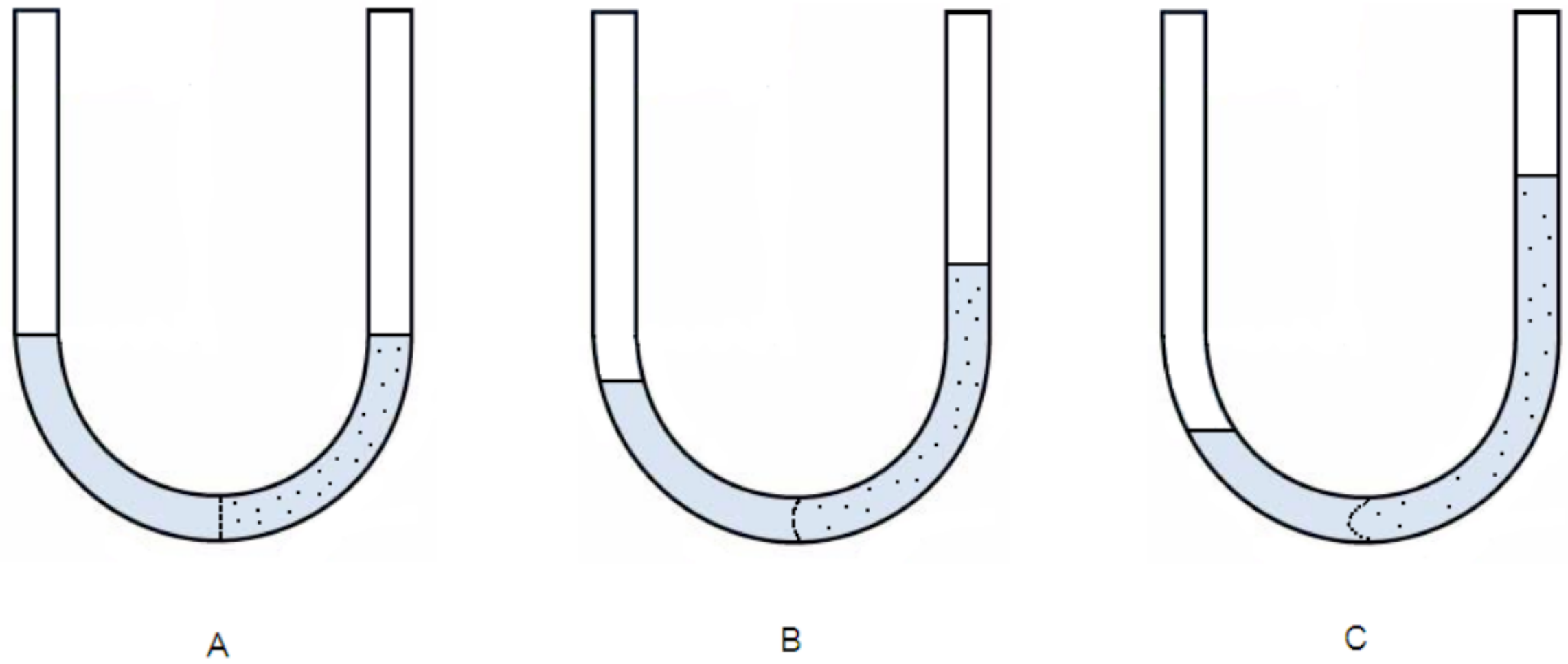

Figure 1 The osmotic process (water in light blue is the osmolyte here and particles on the right side of the tubing are not semipermeable osmolyte here).

(A) In the beginning of the osmosis (the initial instant of the process).

(B) In the middle of osmosis (the process is still going on).

(C) In the end of osmosis (osmotic equilibrium has been reached).

Explanation: At the very beginning (Fig. 1A), although the net osmotic force on the solvent side is maximal, the hydraulic pressures on both sides are the same. Therefore, the membrane does not experience macroscopic elastic deformation. Along with the osmosis, the hydraulic pressure on the solvent side gradually decreases due to the declining water level, but on the contrary the hydraulic pressure on the solution side increase. This pressure difference causes the membrane convex to the solvent side gradually. Finally, when the net osmotic forces become zero, the system arrives at osmotic equilibrium. The pressure difference between both sides, however, reaches the maximum, so the membrane is convex to the solvent side at the maximal extent (Fig. 1C).

## 1.4. Reasons for inability of van't Hoff equation to explain osmotic process

The basic reason for the inapplicability of van't Hoff equation on all solution conditions is in the incomplete concept of van't Hoff's law for the osmotic pressure. It was not abstracted from the dynamic process that fully reveals the nature of osmosis. On the contrary, it was abstracted from the static process by deriving from resisting the osmotic process. The osmotic pressure π was forced into the solution side to impede the osmotic process from solvent (on the left of Fig. 1) and maintain the membrane at the state shown in Fig. 1A. This viewpoint can be further described by van't Hoff osmotic equation, $\pi = \frac{nRT}{v}$ . In this formula, except for the ideal gas constant (R) and the fixed absolute temperature (T) under certain conditions, the osmotic pressure only relates to solute concentration. As a concept to describe the osmotic process, the osmotic pressure formula above does not cover another necessary factor in osmotic process, that is, the strength or force of the pressure, i.e., the metric that changes along the progress of osmotic process. Due to this lack of completeness, no wonder this equation cannot explain osmotic process in some conditions. In contrast to that, the formulas of osmotic force ($\mathrm{f} = P_X \cdot s$, $\mathrm{f} = P \cdot s_X$, or $\mathrm{f} = P_X \cdot S \cdot (1 - k[C_i])$ as described in the section 1.1 above) cover both necessary factors, solute's molar concentration and pressure. Thus, it could explain the osmotic process better.

For the current scientific concepts, if they cannot reflect the true meaning of what they represent for, they need to be improved by some other concepts. Some concepts of van't Hoff's classical osmotic theory cannot explain the related osmotic problems well, and it should be improved by a new osmotic theory.

## 2. RESULTS

## New osmotic pressure formula

The van't Hoff equation(Kiil, 2003)(Prickett et al., 2008)(Rodenburg et al., 2017)(Lopez and Hall, 2020) is generally considered as classical theory because it has been proved mathematically by follower researchers. However, the authors find that there is a far-fetched mistake in the integral substitution which was aimed to mathematically prove the correctness of linear van't Hoff equation. Therefore, the authors derived a new curvilinear osmotic pressure formula based on osmotic process which is able to overcome the limitation of linear van't Hoff equation.

### 2.1 Mistake in mathematical derivation for van't Hoff equation

The mathematical derivation of van't Hoff equation was demonstrated by solving the polynomial equations which are based on multiphase equilibrium of the chemical potential(Wikipedia), as briefly described below:

While in osmotic equilibrium, the chemical potentials on both pure solvent and solution sides balance with chemical potential of gas phase respectively, namely:

$$\mu_A = \mu^\theta + RTlna_A + \int_{p^\theta}^{p_A} v_1 dp \quad (1)$$

$$\mu_B = \mu^\theta + RTlna_B + \int_{p^\theta}^{p_A+\pi} v_1 dp \quad (2)$$

in which, $\mu_A$, $\mu_B$, are the chemical potential on pure solvent side (addressed as A) and solution side (addressed as B), $p_A$, $p_A + \pi$, are pressures on each side, $\pi$ is osmotic pressure, $p^\theta$ is standard pressure, $\mu^\theta$ is standard chemical potential, $v_1$ is molar volume of solvent, $a_A$, $a_B$ are solvent activity on each side.

Since the chemical potential on both sides are equal at equilibrium,

$$\mu_A = \mu_B \quad (3)$$

Substituting equation (1), (2) into equation (3), then

$$\mu^\theta + RTlna_A + \int_{p^\theta}^{p_A} v_1 dp = \mu^\theta + RTlna_B + \int_{p^\theta}^{p_A+\pi} v_1 dp \quad (4)$$

Solving this equation, then we can obtain the van't Hoff equation $\pi = \frac{nRT}{v}$.

There is a logical mistake during the abovementioned derivation process. That is because there is the fundamental difference in the mechanism between osmotic equilibrium and either phase equilibrium or chemical equilibrium. The assumption that chemical potential is equal at equilibrium, which is suitable for phase equilibrium and chemical equilibrium, is not applicable to osmotic equilibrium. To be specific, in osmotic system, there is a functional semi-permeable membrane between the solvent and the solution sides to select specific molecules to pass through. And this membrane is non-existent in phase change or chemical reaction. The equations (1) and (2) above are based on the premise that there is no functional membrane. But from equation (3) to (4), the functional membrane has existed. That is, van't Hoff equation was originally derived on the logic priori of no functional membrane involved, and in the process of derivation, the condition with functional membrane in the middle was manually posed in order to fit the subjective demand. There is one more point, whatever it is for the phase change between solid, liquid and gas or the phase transitions of allotropes such as graphite and diamond, there are changes in energy among atoms or molecules via van der Waals force or hybrid orbital changes. In chemical reaction, there are obviously existing thermal energy changes via old bond breakage and new bond formation. However, osmotic process does not involve similar energy changes. Therefore, regardless of these differences, it was biased to apply the criterion "chemical potential is the same at equilibrium" to the osmotic pressure derivation. It does not mean there is anything wrong with the thermodynamics theory here, it's just that the theory has been used in the wrong place.

In some textbooks, the Taylor series expansion method was borrowed to prove the van't Hoff equation, which is even more wrong. It contains not only the abovementioned mistake, but also the following issues:

1) When applying Taylor series expansion, the equations were approximated more than once, which exposes the inaccuracy from the form.

2) The most important thing is that they omitted the remainders by approximation, which is exactly the component that converts the linear equation into a curvilinear upward one. In other word, it cut off the functional components (which can reveal the exact osmotic change) in order to meet the linear van't Hoff equation.

3) From the Taylor series expansion, it could be seen that only when solution concentration decreases to 0, the approximation will disappear. Even if it is infinitely close to 0, there is bias from the exact value.

From the Taylor expansion series, we can also see, when should we not use approximately equal ($\approx$) but to use all equals ($=$) without omission? Only when the concentration is as small as 0 will it be all equal. But what's the point of being equal when the concentration is 0! No matter how dilute the concentration is, it cannot be zero. This on the other hand proves the fact that when the concentration is 0, it is all equal; once the concentration is slightly higher than 0, even if it is infinitely close to 0 and slightly higher than 0, it has deviated theoretically from the objectively expected curve without cutting by not using "approximately equal to" ($\approx$) but using "full equal to" ($=$). (this fact is in line and consistent with the proof that "van't Hoff equation" is equal to the differential of our proposed new osmotic pressure formula at 0 solute concentration that will be described later in this article).

Therefore, we can say that the method borrowed from the Taylor series expansion cannot prove the correctness of the van't Hoff linear formula, we analysed the errors in its proof, and proved the correctness of the curvilinear feature of proposed new osmotic pressure formula.

The last thing to point out is that the van't Hoff equation was not derived by van't Hoff himself. In 1886, he found that the osmotic pressure of dilute solution can be approximated by applying gas state equation, which was a totally empirical formula. Today's semi empirical and semi theoretical concept of van't Hoff equation was derived by follower scholars. It can be seen that van't Hoff's osmotic pressure formula is actually only a purely empirical formula, and it is inevitable that the purely empirical formula lacking theoretical support has its limitations.

## 2.2 New osmotic pressure formula(Huang and Xie, 2012)

Because the mechanism of osmotic equilibrium is more intuitively reflected in the principle of permeability of membrane, a new osmotic pressure formula can be derived by applying the osmotic law which is based on the core concept of osmotic force, and by following the principle of membrane permeability.

In osmotic equilibrium, the higher part of the hydrostatic pressure on the solution side over the pure solvent side is the osmotic pressure. According to the osmotic force equilibrium formula:

atmospheric pressure · osmotic effective area on the solvent side = (atmospheric pressure + osmotic pressure) · osmotic effective area on the solution side

The abovementioned formula was established on the ideal state that the temperature of solvent is equal to its melting temperature ($T_0$=273.6 K for water). Under the actual condition that temperature is higher than the melting temperature of solvent, the osmotic pressure on the solution side would be enhanced due to increasing solvent mobility and higher osmotic effective area on the solvent side. The amount of the enhancement is determined by the ratio of absolute temperature at actual state (T) to the absolute temperature at melting point of solvent ($T_0$), i.e. T/ $T_0$. Then, the new osmotic pressure formula is derived on the basis of osmotic force equilibrium formula as:

$$\text{osmotic pressure} = \left(\frac{\textit{atmospheric pressure} \cdot \text{osmotic effective area on the solvent side}}{\text{osmotic effective area on the solution side}} - \textit{atmospheric pressure}\right) \cdot \frac{T}{T_0}$$

Addressing the osmotic pressure as $\pi$, atmospheric pressure as $p$, osmotic effective area on the solvent side as $s_a$, osmotic effective area on the solution side as $s_b$, and the molar concentration of solution as $[C_i]$. Known that the osmotic effective area on the solvent side is equal to the total membrane area S, and the osmotic effective area on the solution side $s_b = (1 - k[C_i]) \cdot S$, the new osmotic pressure formula could be derived and expressed as:

$$\pi = \left(\frac{p \cdot s_a}{s_b} - p\right) \cdot \frac{T}{T_0} = \left[\frac{P(s_a - s_b)}{s_b}\right] \cdot \frac{T}{T_0} = \frac{p[1-(1-k[C_i])]}{(1-k[C_i])} \cdot \frac{T}{T_0} = \frac{pk[C_i]}{1-k[C_i]} \cdot \frac{T}{T_0}$$

Then, the new formula is $\pi = \frac{pk[C_i]}{1-k[C_i]} \cdot \frac{T}{T_0}$. (5)

## 2.3 The features of new osmotic pressure formula:

The first-order derivative of the new osmotic pressure formula with respect to molar concentration is:

$$\pi' = \frac{d\,\pi}{d[C_i]} = \frac{d\left(\frac{pk[C_i]}{1-k[C_i]} \cdot \frac{T}{T_0}\right)}{d[C_i]} = \frac{pk(1-k[C_i]) - pk[C_i](-k)}{(1-k[C_i])^2} \cdot \frac{T}{T_0} = \frac{pk}{(1-k[C_i])^2} \cdot \frac{T}{T_0}$$

Then the second-order derivative is:

$$\pi'' = \frac{d}{d[C_i]}\left(\frac{d\,\pi}{d[C_i]}\right) = \frac{-pk \cdot 2(1-k[C_i])(-k)}{(1-k[C_i])^4} \cdot \frac{T}{T_0} = \frac{2pk^2}{(1-k[C_i])^3} \cdot \frac{T}{T_0}$$

$p, k, T, T_0$ are positive number, so $\pi' > 0$ and $\pi'' > 0$ within the function domain [0, 1/k). It not only means that the original function (i.e. the new osmotic pressure formula) is one-way incremental and the function curve is concave upward, but also that $\pi'$ is one-way incremental (proved by $\pi'' > 0$). Thus, with the increase of solution concentration, the osmotic pressure increases rapidly.

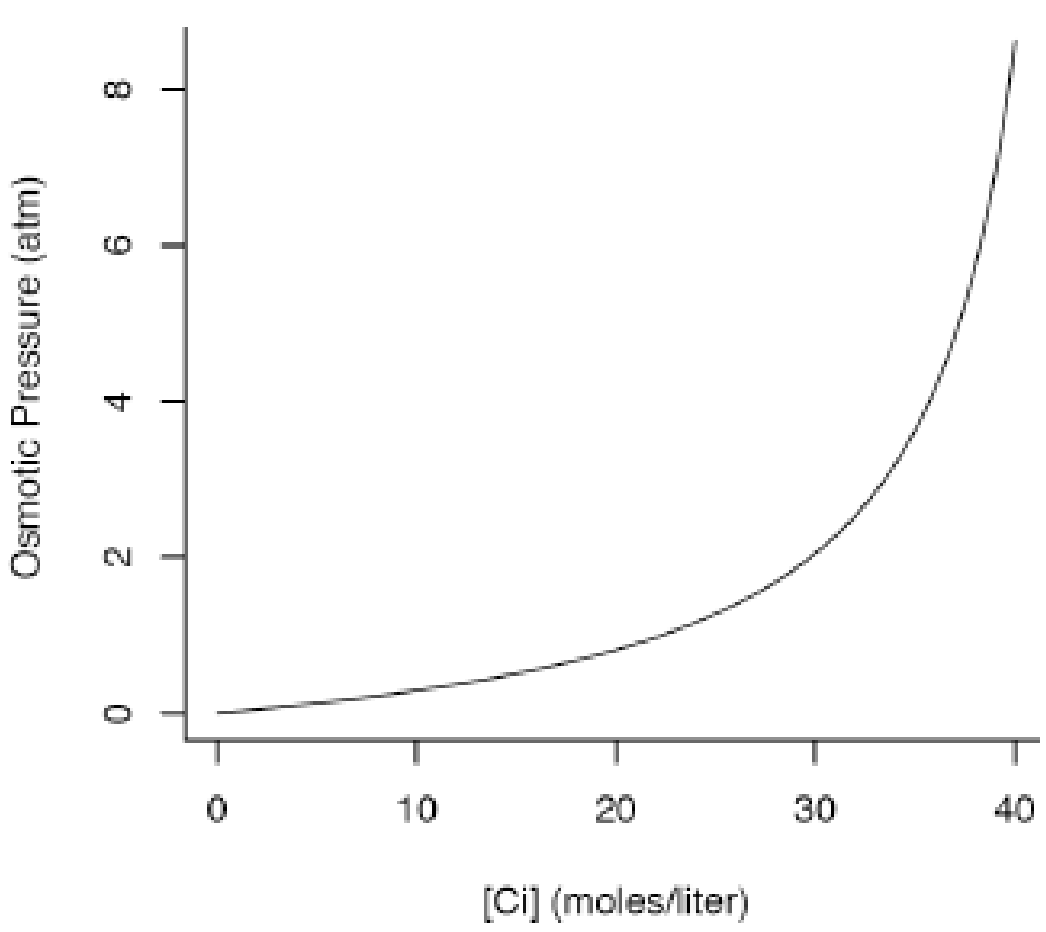

Figure 2 The plot for the change of osmotic pressure (***π***) versus the change of solute concentration [Ci] according to the formula in equation (5) if *p*=1 atm, *k*=0.0224 (derived from the R value for dilute ideal solution), and $\frac{T}{T_0}$=1. When the [Ci] is low (e.g., less than 10 moles/liter), the curve is almost linear; the linear curve in this part is close to the linear equation for van't Hoff osmotic pressure formula, i.e., π = (*n/V*)RT or π = [Ci]RT.

**This curve contains three levels of reactions:**

(1) first level reaction: molar concentration ($[C_i]$) is very low, i.e. $\mathrm{k}[C_i] \ll 1$, then $\pi \approx \mathrm{pk}[C_i]$. The osmotic pressure is proportional to molar concentration, and the curve is linear in this segment.

(2) mixed level reaction: molar concentration ($[C_i]$) is relatively high. osmotic pressure increase swiftly along with the increasing molar concentration. The curve is concave upward in this segment.

(3) zero level reaction: molar concentration ($[C_i]$) is maximal and the solution is saturated. The numerator $\mathrm{pk}[C_i]$ of the formula $\pi = \frac{pk[C_i]}{1-k[C_i]} \cdot \frac{T}{T_0}$ reaches the maximum, while the denominator $1 - k[C_i]$ reaches the minimum. The curve gets the peak point.

## 2.4 Proof of the experimental data on the new osmotic pressure formula

Table 1. This table shows the osmotic pressures of NaCl solution at three different concentrations (25℃)

| component | concentration (mg/L) | osmotic pressure (MPa) |
| --- | --- | --- |
| NaCl | 1000 | 0.078 |
| NaCl | 2000 | 0.16 |
| NaCl | 5000 | 0.46 |

(Data were from (Tang and Dai, 2008) and (Zhu, 2013))

It can be seen that the osmotic pressure of NaCl solution is not linearly proportional to the increasing of molar concentration. That is, when the ratio of the concentration is “1:2:5”, the ratio of osmotic pressure is “1 : 2.05 : 5.89”, and it can be seen that the plot of osmotic pressure vs concentration is not a straight line but a curve with upward bending tendency, which shows the consistency with the new osmotic pressure formula theory resulting a plot showing concave upward curve. Similar variation trends could be seen from $CuSO_4$, $BaCl_2$, $Ca(NO_3)_2$, $CaCl_2$, $NH_4Cl$ solutions (Xia, 2008). This phenomenon is consistent with the new osmotic pressure theory.

## 2.5 Experimental Evidence

We used “WP4C Dewpoint PotentiaMeter” manufactured by Decagon Device Inc.(WP4-C Dew Point PotentiaMeter. Operator’s manual, version 1, Decagon Devices, Inc., 2003)(Petry and Jiang, 2003)(Maček et al., 2013) to measure the osmotic pressure on NaCl solutions with different molar concentrations. We utilized the classical van’t Hoff equation and the proposed new osmotic pressure formula to calculate the osmotic pressure with known concentration of NaCl solutions, then compare these data with the experimental data obtained by direct measure with Decagon’s PotentiaMeter. It was discovered from these data that the calculated data via proposed new formula by us have better correlation and proportion to the actual measured data when the solution concentration was increased higher.

The Dewpoint PotentiaMeter, also known as a chilled-mirror hygrometer, measures dewpoint and temperature very accurately in a closed space above the soil sample. The sealed chamber has a mirror and condensation detector with precise temperature control. At equilibrium, the relative humidity of the air in the chamber is the same as the relative humidity of the soil sample. When the first condensation appears on the mirror, the vapor pressure is measured and total suction can be calculated(WP4-C Dew Point PotentiaMeter. Operator’s manual, version 1, Decagon Devices, Inc., 2003)(Petry and Jiang, 2003)(Maček et al., 2013).

The measured data and the theoretical values calculated by using both van’t Hoff equation and our proposed new osmotic pressure formula are shown in the following table.

Table 2. Osmotic suction measurement of salt solutions

| NaCl concentration (mol / L) | van’t Hoff equation (MPa) | New osmotic pressure formula (MPa) | measured suction by PotentiaMeter (MPa) |
|---|---|---|---|
| 0.5 | 2.44 | 2.48 | 1.95 |
| 3 | 14.616 | 16.83 | 16.04 |
| 4.5 | 21.924 | 27.37 | 28.2 |

# 3. DISCUSSION

## Understanding the van’t Hoff equation and new osmotic pressure formula by comparison

### 3.1 van't Hoff equation is the differential of the new osmotic pressure formula at 0

The derivative of new osmotic pressure formula at the independent variable $[C_i] = 0$ is:

$$\mathrm{f}'(0) = \pi'|_{[C_i]=0} = \frac{d\,\pi}{d[C_i]}|_{[C_i]=0} = \frac{pk(1-k[0]) - pk[0](-k)}{(1-k[0])^2} \cdot \frac{T}{T_0} = \frac{pkT}{T_0}$$

That is, $$\pi'|_{[C_i]=0} = \frac{pkT}{T_0} \quad (6)$$

Because the new osmotic pressure formula was strictly derived from the actual osmotic equilibrium and its curve is concave upward in correspondence to experimental data as seen in Table 2, its validity was proved by both theoretical and empirical truth. On the other hand, the applicability of van’t Hoff equation in very dilute solution indicates that these two formulas are the same at dilute solution state from a numerical perspective, that is:

$$\frac{pk[C_i]}{1-k[C_i]} \cdot \frac{T}{T_0} = \frac{n}{v}RT$$

In which $\frac{n}{v}$ is an idiomatic usage to describe molar concentration, and it is the same as $[C_i]$. Thus,

$$\frac{pk}{1-k[C_i]} \cdot \frac{T}{T_0} = RT$$

$$pk \cdot \frac{T}{T_0} = RT(1-k[C_i])$$

and $lim_{[C_i]\to 0}\mathrm{RT}(1-k[C_i]) = RT$

so $pk \cdot \frac{T}{T_0} = RT$ when $[C_i]$ is close to 0. (7)

According to equations (6), (7) above, then

$$\pi'|_{[C_i]=0} = RT \quad (8)$$

Equation (8) illustrates that the derivative of the new osmotic pressure formula at $[C_i] = 0$ is the slope of van't Hoff equation, RT. Therefore, van't Hoff equation is the differential of the new osmotic pressure formula at 0 solute concentration.

Now the value and unit of constant k in new osmotic pressure formula could be determined by the differential relationship between van't Hoff equation and new osmotic pressure formula.

Since $$\mathrm{R} = \frac{pV_{m,0}}{T_0} = 8.314\, J/(mol \cdot K) \quad (9)$$

and $$pk \cdot \frac{T}{T_0} = \mathrm{RT} = \frac{pV_{m,0}T}{T_0} \quad (10)$$

Substituting equation (9) into equation (10), then

$$k = V_{m,0} = 22.4 \times 10^{-3}\, m^3/mol \quad (11)$$

Therefore, if van't Hoff equation is assumed to be correct at ideal dilute solution state, the k in new osmotic pressure formula should be $22.4 \times 10^{-3}\, m^3/mol$.

There are two points to explain: 1) after calculation with new osmotic pressure formula and unit conversion with this k value, the unit of osmotic pressure $\pi$ calculated by new formula is still Pa, 2) k value experiences slight variation due to the different structure of different solute molecules, so the actual k used in application should be based on the experimentally measured value.

From the above, the difference in format between the two formulas can be further derived.

Due to equation (7), the van't Hoff equation can be expressed as:

$$\pi = \frac{n}{v}pk \cdot \frac{T}{T_0}$$

Because $\frac{n}{v} = [C_i]$, so it could be rearranged as:

$$\pi = pk[C_i] \cdot \frac{T}{T_0} \tag{12}$$

From the equation (12), it can be seen that the difference between van't Hoff equation and new osmotic pressure formula in equation (5) lies on the denominator, i.e. $1 - k[C_i]$. And this difference well explains the discrepancy in function graph as well as the similarity at very dilute solution.

## 3.2 The reason for the error of van't Hoff equation

The van't Hoff equation is similar in form to the atmospheric pressure formula of ideal gas. However, the effect of solute molecules on osmotic pressure is quite different form the effect of gas molecules on atmospheric pressure. For ideal gas, the atmospheric pressure is linearly proportional to the independent variable, gas density (formula for non-ideal gas can be modified on this basis), which has been proved by experiments. However, osmotic pressure does not have the linear relationship with molar concentration of solution. It is not only related to the osmotic ineffective area but also to osmotic effective area, which are both influenced by solution concentration but play opposite roles with sum of 1. To be specific, osmotic ineffective area enhances osmotic pressure, while osmotic effective area decreases osmotic pressure. When the molar concentration of solution increases, the osmotic ineffective area would increase and the osmotic effective area would decrease accordingly. Both processes enhance the osmotic pressure, thus it results in the concave upward form of the osmotic pressure curve rather than a linear relationship as van't Hoff equation depicts. Some people may argue that the bias in van't Hoff equation can be modified by introducing activity coefficient. However, it is not the case. If the activity was the factor which affected the accuracy of ideal formula, the formula would be rational to be modified by activity coefficient. The linear van't Hoff equation was originally a fundamentally theoretical error. In this case, the correction with the activity coefficient alone would not help.

This standpoint can be further demonstrated by osmotic equilibrium formula:

$$\begin{aligned}&\text{atmospheric pressure} \cdot \text{osmotic effective area on solvent side}\\ &\qquad = (\text{atmospheric pressure} + \text{osmotic pressure}) \cdot \text{osmotic effective area on the solution side} \cdot \frac{T}{T_0}\end{aligned}$$

Because the left-hand side of the equation is constant and the $\frac{T}{T_0}$ is constant at certain condition, osmotic effective area on the solution side will decrease when solution concentration increases. To reach equilibrium, (atmospheric pressure + osmotic pressure) should increase proportionally according to the increase of concentration. Since the atmospheric pressure is constant, it could only be achieved by increasing osmotic pressure. Moreover, with the increase of solution concentration, osmotic pressure would be required to increase faster, as reflected by the concave upward curve.

## 3.3 Quadrant rules for the new osmotic pressure formula in reality

Theoretical formulas derived under ideal conditions all require modifications in real application in unideal conditions, and the new osmotic pressure formula is no exception. In practice, the formula should be calibrated by activity coefficient. Since the activity coefficient changes along with the variation of solution concentration, the quadrant for new osmotic pressure formula has the following regular changes:

1) within a certain concentration range, if the activity coefficient increases with the increasing concentration, the upward steepness of the plot in this quadrant would increase.

2) within a certain concentration range, if the activity coefficient is constant, the shape of the plot in this quadrant is related to fixed activity (ideal solution has maximum fixed activity, and the shape of the plot for conventional osmotic pressure in this quadrant belongs to this category).

3) within a certain concentration range, if the activity coefficient decreases with the increasing concentration, the upward steepness of the plot in this quadrant would decrease, even with a concave upward trend.

Based on the new osmotic pressure formula, the whole plotting curve of osmotic pressure is the concatenation of the plots in different quadrants formed at different concentration range with different activity (coefficient) variation.

## 4. Conclusions

The history of science development tells us that when people were temporarily unable to explore the inherent logical nature of a certain scientific phenomenon, the issues could only be dealt with less rigorous or limited empirical methods in order to initially summarize and solve the needs of the time. This is the case with the van't Hoff equation proposed more than a hundred years ago, so it is inevitable that it has imperfect limitations. However, as people gradually grasp the inherent logical nature of the osmosis phenomenon during practical practice, the limited empirical formula can be improved and optimized by the relatively scientific and rigorously theoretical new formula proposed in this article. This is the inevitable progress due to continuous development of scientific knowledge and understanding.

**Keywords: osmotic force; osmotic pressure; semipermeable membrane; osmotic effective area; van't Hoff equation**

## Acknowledgements:

This research was supported by the National Institutes of Health/National Institute on Minority Health and Health Disparities Grant Number:1U54MD015929-01, through the RCMI-Center for Health Disparities Research at Jackson State University.

## References:

Huang, H.-C., and R. Xie. 2012. New Osmosis Law and Theory: the New Formula that Replaces van't Hoff Osmotic Pressure Equation. *arXiv*. 1201.0912.

Kiil, F. 2003. Kinetic model of osmosis through semipermeable and solute-permeable membranes. *Acta Physiol. Scand.* 177:107–117. doi:10.1046/j.1365-201X.2003.01062.x.

Lopez, M.J., and C.A. Hall. 2020. Physiology, Osmosis. StatPearls Publishing.

Maček, M., J. Smolar, and P. Ana. 2013. Extension of measurement range of dew-point potentiometer and evaporation method. 18th International Conference on Soil Mechanics and Geotechnical Engineering (At: Paris).

Petry, T.M., and C.-P. Jiang. 2003. EVALUATION AND UTILIZATION OF THE WP4 DEWPOINT POTENTIAMETER PHASE I AND II. *Cent. Infrastruct. Eng. Stud. Progr. Univ. Missouri - Roll.*

Prickett, R.C., J.A.W. Elliott, S. Hakda, and L.E. McGann. 2008. A non-ideal replacement for the Boyle van't Hoff equation. *Cryobiology*. 57:130–136. doi:10.1016/j.cryobiol.2008.07.002.

Rodenburg, J., M. Dijkstra, and R. Van Roij. 2017. Van't Hoff's law for active suspensions: The role of the solvent chemical potential. *Soft Matter*. 13:8957–8963. doi:10.1039/c7sm01432e.

Tang, S., and Y. Dai eds. . 2008. Water Treatment Engineer Handbook, Table 2-14-7. 1st ed. Chemical Industry Press. Table 2-14-7 pp.

Wikipedia. Osmotic Pressure.

WP4-C Dew Point PotentiaMeter. Operator's manual, version 1, Decagon Devices, Inc. 2003.

Xia, Y. ed. . 2008. Chemistry Laboratory Manual. 2nd ed. Chemical Industry Press. 997 pp.

Zhu, Y. ed. . 2013. Water treatment technology equipment design manual, Table 8-11. 1st ed. China Architecture & Building Press.